\documentclass[sigconf]{acmart}

\usepackage{mdframed}

\AtBeginDocument{%
  }

\setcopyright{acmlicensed}
\copyrightyear{2024}
\acmYear{2024}
\acmDOI{XXXXXXX.XXXXXXX}

\acmConference[Conference acronym 'XX]{Make sure to enter the correct
  conference title from your rights confirmation emai}{June 03--05,
  2018}{Woodstock, NY}

\acmISBN{978-1-4503-XXXX-X/18/06}

\usepackage{multirow}
\usepackage{amsmath}
\usepackage{color}
\usepackage{tikz}
\usepackage{bbm}
\begin{document}

\title{Laser: Parameter-Efficient LLM Bi-Tuning for  Sequential Recommendation with Collaborative Information}

\author{Xinyu Zhang}
\email{xyzhang0105@gmail.com}
\affiliation{%
  \institution{Beijing Institute of Technology}
  \state{Beijing}
  \country{China}
}

\author{Linmei Hu}
\email{hulinmei@bit.edu.cn}
\affiliation{%
  \institution{Beijing Institute of Technology}
  \state{Beijing}
  \country{China}
}

\author{Luhao Zhang}
\email{zhangluhao@bit.edu.cn}
\affiliation{%
  \institution{Beijing Institute of Technology}
  \state{Beijing}
  \country{China}
}

\author{Dandan Song}
\email{sdd@bit.edu.cn}
\affiliation{%
  \institution{Beijing Institute of Technology}
  \state{Beijing}
  \country{China}
}

\author{Heyan Huang}
\email{hhy63@bit.edu.cn}
\affiliation{%
  \institution{Beijing Institute of Technology}
  \state{Beijing}
  \country{China}
}

\author{Liqiang Nie}
\email{nieliqiang@gmail.com}
\affiliation{%
  \institution{Harbin Institute of Technology}
  \state{Shenzhen}
  \country{China}
}

\begin{abstract}

Sequential recommender systems are essential for discerning user preferences from historical interactions and facilitating targeted recommendations. Conventional techniques rely solely on item IDs for sequence modeling, overlooking the wealth of semantic data in item descriptions, which can lead to subpar performance.
Recent innovations employing Large Language Models (LLMs) have advanced the field by encoding item semantics, yet they often necessitate substantial parameter tuning and are resource-demanding. Moreover, these works typically integrate  ID-based collaborative signals into LLMs via a simple unified linear projection, which fails to consider the diverse characteristics of different types of users and thus diminishes the recommendation accuracy.

In this paper, we propose a parameter-efficient \underline{La}rge Language Model Bi-Tuning framework for \underline{se}quential \underline{r}ecommendation with collaborative information ({Laser}). Specifically,  Bi-Tuning works by inserting trainable virtual tokens at both the prefix and suffix of the input sequence and freezing the  LLM parameters, thus optimizing the LLM for the sequential recommendation. In our Laser, the prefix is utilized to incorporate user-item collaborative information and adapt the LLM to the recommendation task, while the suffix converts the output embeddings of the LLM from the language space to the recommendation space for the follow-up item recommendation.
Furthermore, to capture the characteristics of different types of users when integrating the collaborative information via the prefix, we introduce M-Former, a lightweight MoE-based querying transformer that uses a set of query experts to integrate diverse user-specific collaborative information encoded by frozen ID-based sequential recommender systems, significantly improving the accuracy of recommendations.

Extensive experiments on real-world datasets demonstrate 
that Laser can parameter-efficiently adapt LLMs to effective recommender systems, significantly outperforming state-of-the-art methods. 
\end{abstract}

\begin{CCSXML}
<ccs2012>
<concept>
<concept_id>10002951.10003317.10003347.10003350</concept_id>
<concept_desc>Information systems~Recommender systems</concept_desc>
<concept_significance>500</concept_significance>
</concept>
</ccs2012>
\end{CCSXML}

\ccsdesc[500]{Information systems~Recommender systems}

\keywords{Sequential recommendation, large language model, parameter-efficient fine-tuning, MoE, collaborative information}

\received{20 February 2007}
\received[revised]{12 March 2009}
\received[accepted]{5 June 2009}

\maketitle

\section{Introduction}

Sequential recommender systems are designed to learn effective representations of users' interests based on their past interactions and to suggest future items that match users' needs. Due to their abilities to capture the dynamic nature of user preferences and their effectiveness in enhancing user satisfaction, sequential recommender systems are widely applied in various scenarios such as e-commerce, streaming services, and social media platforms \cite{ DBLP:conf/kdd/ChongLZZLLW0L23, DBLP:conf/kdd/ZhangWC023, DBLP:conf/kdd/GholamiMA22, DBLP:journals/corr/HidasiKBT15}.

In traditional sequential recommender systems, items are predominantly represented by unique IDs.
To obtain effective ID embeddings {based on the user interaction sequence}, a variety of methods are employed, including Markov Chains \cite{DBLP:conf/icdm/HeM16, DBLP:conf/www/RendleFS10}, RNN/CNN models \cite{DBLP:journals/corr/HidasiKBT15, DBLP:conf/cikm/LiRCRLM17, DBLP:conf/wsdm/TangW18, DBLP:conf/wsdm/YuanKAJ019}, and self-attentive models \cite{DBLP:conf/icdm/KangM18, DBLP:conf/wsdm/LiWM20, DBLP:conf/cikm/SunLWPLOJ19}. While ID-based methods are promising in capturing latent associations between users and items, they fail to consider the rich semantic information contained in the textual descriptions of items (e.g., item title),  resulting in suboptimal performance. 
To solve this issue, efforts have been made to encode item semantic information with language models  \cite{DBLP:conf/kdd/HouMZLDW22, DBLP:conf/kdd/LiWLFSSM23}. However, previous works mainly focus on small or medium-sized language models, which exhibit limited performance.

\begin{figure}[t] %H为当前位置，!htb为忽略美学标准，htbp为浮动图形
    \centering %图片居中
     \vspace{0.1in}
    \includegraphics[width=0.5\textwidth]{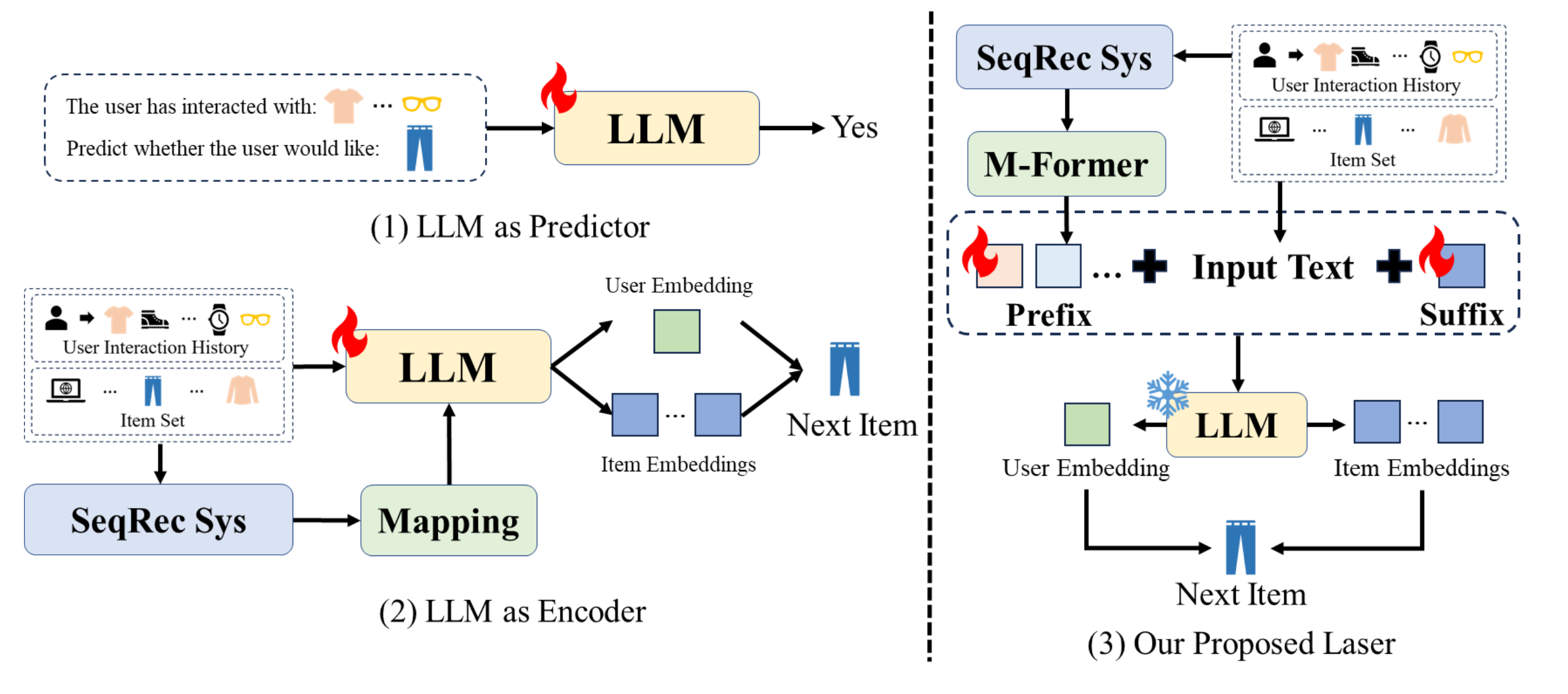} %插入图片，[]中设置图片大小，{}中是图片文件名
     % \vspace{-0.、2in}
    \caption{\textcolor{black}{Comparison of existing methods and our proposed Laser.}} %最终文档中希望显示的图片标题
     \vspace{-0.22in}
    \label{butterfly} %用于文内引用的标签
\end{figure}

Recently, Large Language Models (LLMs) have made significant progress in language understanding \cite{DBLP:journals/corr/abs-2307-16645, DBLP:journals/corr/abs-2401-17043, DBLP:journals/corr/abs-2307-09288, DBLP:journals/corr/abs-2312-17617}. It is a promising way to harness the powerful semantic information modeling capabilities of LLMs pre-trained on extensive text corpora to capture the semantic information of items. 
As shown in Figure \ref{butterfly}, existing works integrate LLMs into recommendation tasks in two main paradigms. The first paradigm is to use LLMs to directly recommend in the form of natural language. These works design special prompts \cite{DBLP:journals/corr/abs-2308-16505, DBLP:journals/corr/abs-2304-10149, DBLP:journals/corr/abs-2308-14296} or use supervised fine-tuning \cite{DBLP:journals/corr/abs-2308-08434, DBLP:conf/recsys/BaoZZWF023, DBLP:journals/corr/abs-2305-07001} to get LLMs to answer the given recommendation questions. However, this paradigm can only determine the recommendation for one item at a time and the frequency of LLM utilization increases linearly with the number of candidate items. Thus, these methods tend to be used only in the reranking phase
, which contains only dozens of candidate items \cite{DBLP:conf/ecir/HouZLLXMZ24, DBLP:journals/corr/abs-2305-07001}. 
The second paradigm is to use LLMs as encoders 
to provide item/user embeddings for similarity comparison and next item prediction.
% and recommend based on embedding similarity. 
As shown in Figure \ref{butterfly}, given the user interaction history represented in natural language, these works use LLMs to encode each token in the input text and then
perform various pooling strategies on 
the output token embeddings to derive the user embedding \cite{DBLP:journals/corr/abs-2305-11700, DBLP:conf/sigir/WuWQ021, DBLP:journals/corr/abs-2403-01744}. Although these works are promising, they typically necessitate the training of extensive parameters, demanding considerable computational resources. Furthermore, these works struggle to effectively incorporate ID-based collaborative information into LLMs, which affects the effectiveness of recommendations. Although efforts have been made to use simple linear projections to map the collaborative embeddings into the language space of LLMs \cite{DBLP:journals/corr/abs-2310-20487, DBLP:journals/corr/abs-2310-19488}, these methods fail
to consider the diverse characteristics of various types of users, potentially resulting in inferior recommendation results.

% 加一张图
%结束环境

To address the above issues, in this paper, we propose a parameter-efficient \underline{La}rge Language Model Bi-Tuning framework for \underline{se}quential \underline{r}ecommendation (Laser), which also effectively integrates collaborative information by capturing the characteristics of different types of users through an MoE-based querying transformer.  In particular, to efficiently adapt LLMs to effective sequential recommender systems that can provide high-quality item/user embeddings, we design a parameter-efficient bidirectional LLM fine-tuning method, named Bi-Tuning.  In Bi-Tuning, we freeze LLMs' parameters and tailor them for recommendation tasks by optimizing the trainable virtual tokens added at the prefix and suffix of the input text, which largely reduces the scale of parameters that require training. The prefix tokens can be utilized to incorporate collaborative information and are responsible for adapting LLMs to recommendation tasks, while the appended single suffix token aims to convert the output of LLMs from the language space to the recommendation space for following embedding similarity comparison and next item recommendation. In addition, to effectively incorporate the collaborative information via the prefix for accurate recommendation, we present M-Former, a lightweight MoE (Mixture of Experts) based querying transformer that employs a set of trainable query experts to capture the diverse characteristics of user-specific collaborative information encoded by frozen ID-based sequential recommender systems.  Experimental results on real-world datasets across different domains show that our method outperforms state-of-the-art baselines.
In summary, our main contributions can be summarized as follows:

\begin{itemize}
\item We propose a parameter-efficient Large Language Model Bi-Tuning framework for sequential recommendation, named Laser, which can effectively adapt LLMs to sequential recommender systems in a parameter-efficient way.
\item In our Laser, to effectively incorporate the collaborative information into LLMs for more accurate recommendation, we design M-Former, a lightweight MoE-based querying transformer that employs a set of query experts to capture the characteristics of user-specific collaborative information encoded by frozen ID-based sequential recommender systems.
\item Extensive experiments on real-world datasets demonstrate that our proposed Laser significantly outperforms state-of-the-art methods.
\end{itemize}

% 加一张图
\begin{figure*}[t!] %H为当前位置，!htb为忽略美学标准，htbp为浮动图形
    \centering %图片居中
    \includegraphics[width=0.68 \textwidth]{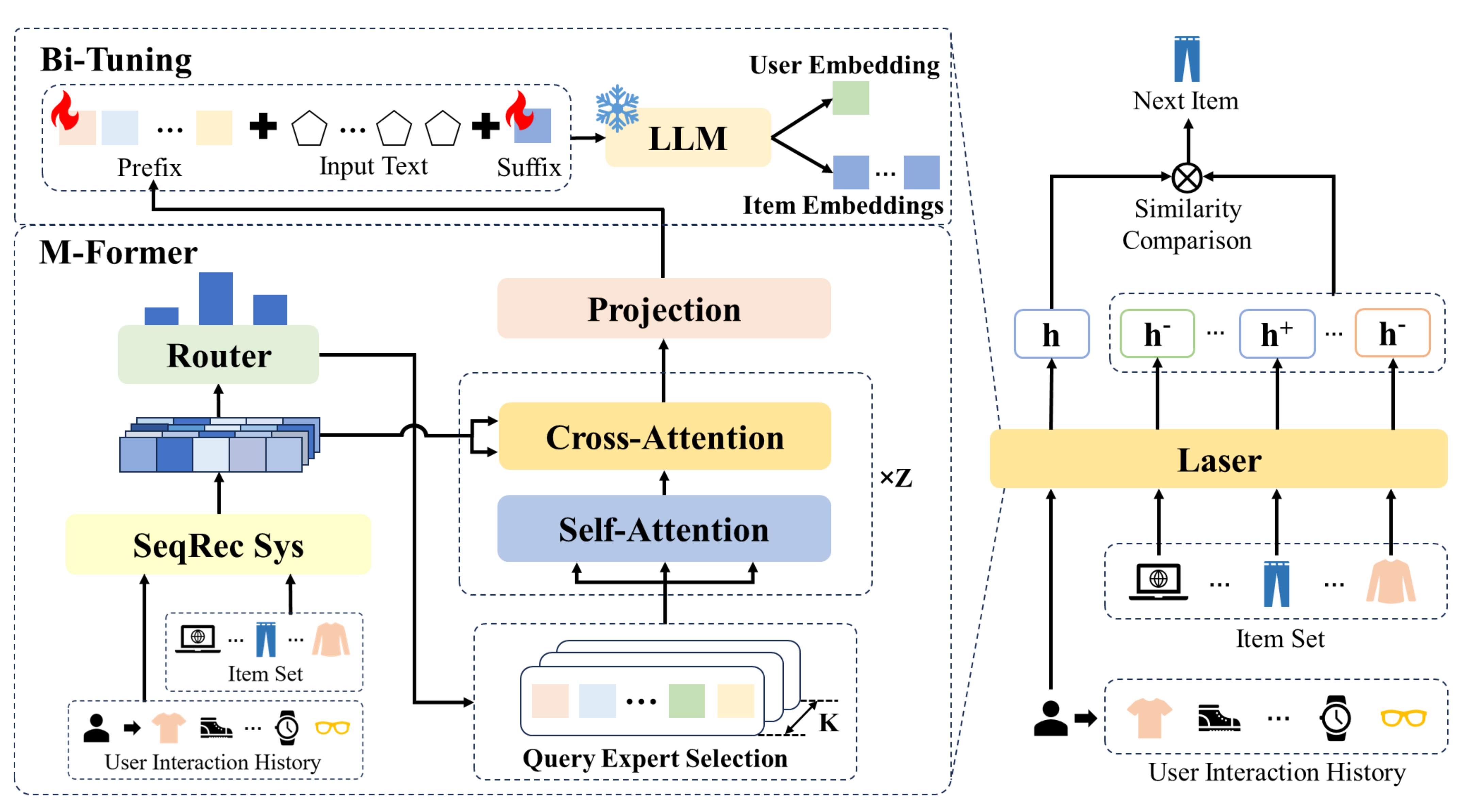} %插入图片，[]中设置图片大小，{}中是图片文件名
     \vspace{-0.1in}
    \caption{\textcolor{black}{The overview of our proposed Laser.}} %最终文档中希望显示的图片标题
    \vspace{-0.05in}
    \label{model} %用于文内引用的标签
\end{figure*}%结束环境

\section{Problem Formulation}
In the setting of sequential recommendation, we are given a user set $\mathcal{U}$ and an item set $\mathcal{I}$. Each user $u\in\mathcal{U}$ is associated with a temporally ordered sequence of his/her historical interacted items, denoted as ${S}_u=\{i_1, i_2,\dots, i_{N}\}$, where $N$ is the length of ${S}_u$ and $i\in\mathcal{I}$. Based on ${S}_u$, sequential recommender systems are used to predict the next item $i_{N+1}$ that user $u$ is most likely to interact with. 

In traditional ID-based sequential recommender systems, each item $i$ is associated with a unique item ID ${id}_i$, and the ID sequence ${ID}_u=\{{id}_{i_1}, {id}_{i_2},\dots,{id}_{i_{N}}\}$ is used as input of the model to predict the next item ID $id_{i_{{N}+1}}$. Differently, in this work, in addition to the item id sequence ${ID}_u$, 
we also utilize the semantic information of items, including the attributes such as title, category, and brand.
 Formally, the attributes of an item $i$ can be represented as ${D}_i=\{(k_1, v_1), (k_2, v_2), \dots, (k_M, v_M)\}$, where $k$ is the attribute name  (e.g., ``title'', ``category'', and ``brand''),  $v$ is the corresponding value  and $M$ is the number of attributes.
Given the user interaction sequence ${S}_u=\{i_1, i_2,\dots, i_{N}\}$, {we use a template to organize the corresponding item attribute sequence $D_u=\{D_{i_1}, D_{i_2}, ..., D_{i_N}\}$ into a complete and coherent text $T_u=\{t_1, t_2, ..., t_W\}$ (detailed in Section \ref{prompt}}), where $W$ is the text length. 
Then, $T_u$ will be taken as the input of LLMs for next item prediction. 

\section{Methodology}
In this section, we detail our proposed Large Language Model Bi-
Tuning framework for sequential recommendation, Laser. 
As illustrated in Figure \ref{model},  a parameter-efficient LLM Bi-Tuning method is presented to adapt LLMs for the sequential recommendation. Moreover, a lightweight MoE-based querying transformer, M-Former, is designed to effectively integrate collaborative information into LLMs while capturing the characteristics of users of different types with the MoE strategy.

\subsection{LLM Bi-Tuning for Sequential Recommendation}

{In the following, we first describe how to organize the user interaction history $D_u=\{D_{i_1}, D_{i_2}, ..., D_{i_N}\}$  into a coherent text $T_u=\{t_1, t_2, ..., t_W\}$, which is taken as the input of LLMs for the sequential recommendation, and then introduce how to adapt  LLMs for the recommendation task by the proposed parameter-efficient Bi-Tuning method.}
\subsubsection{Input Text Formulation}
\label{prompt}
\textcolor{black}{In this work, we utilize a unified template to organize the user interaction history $D_u$ into the input text  $T_u$ of LLMs for recommendation. For example, given the user who has browsed ``Kaytee Aspen Bedding Bag'', ``Guitar A-Frame SupportS'', ..., and ``KONG Wubba Dog Toy'', we can formulate the corresponding input text $T_u$ as follows.}
\vspace{2mm}
\begin{mdframed}[backgroundcolor=black!15] 
\vspace{1.5mm}
\noindent You are an intelligent recommendation assistant. Please summarize the user's characteristics into a \textit{single token} based on the browsing history.
In chronological order, the user has browsed the following items:

\vspace{1mm}

 \noindent $>>$ 1. {Kaytee Aspen Bedding Bag} ({brand: Kaytee}, {category: Kaytee})\\
\noindent $>>$ 2. Guitar A-Frame Support (brand: Sageworks, category: Sageworks)\\
...\\
\noindent $>>$ 4. KONG Wubba Dog Toy (brand: KONG, category: KONG)
\end{mdframed}
\vspace{3mm}

\noindent
\textcolor{black}{
With the designed template, LLMs can follow the instruction in the input text to summarize the user browsing history into a single token (i.e., the suffix appended to the end of the input text), whose corresponding output embedding can be taken as the user embedding $\mathbf{h}_u$ for similarity comparison with the item embeddings and the next item recommendation. To obtain the item embeddings, we also use the above same template. Particularly,  
for a specific item $i$, we treat it as a special user interaction history that contains only this one item. Therefore, we use the same template to formulate the input text of the LLM and take the output appended suffix embedding as the item embedding $\mathbf{h}_i$. In this way, we can obtain the embedding of each item $i$ in the item set $\mathcal{I}$, and the original user-item similarity comparison used for recommendation can be regarded as a special kind of user-user similarity comparison. 
The advantage of this is that a unified template could minimize the impact of hard templates on the performance of LLMs \cite{DBLP:conf/acl/LiuJFTDY022, DBLP:conf/emnlp/LesterAC21}. Detailed experiments in Section \ref{prompt_template_exp} further prove the validity of our unified template.
 }

\subsubsection{Parameter-Efficient Bi-Tuning}

Existing works have shown the powerful capabilities of LLMs in bolstering the sequential recommendation  \cite{DBLP:journals/corr/abs-2308-08434, DBLP:conf/recsys/BaoZZWF023, DBLP:journals/corr/abs-2305-07001}. However, they still face two main challenges:
(1) how to adapt LLMs for recommendation tasks in a parameter-efficient way, and (2) how to effectively transform the output of LLMs from the language space to the recommendation space for the following item recommendation. To solve these two challenges,  as shown in Figure \ref{model}, we propose a parameter-efficient LLM Bi-Tuning method that adapts LLMs through the trainable \emph{prefix} and \emph{suffix}. 

Formally, given the input text $T_u = \{t_1, t_2, ..., t_W\}$, it will be expanded with the trainable prefix and suffix:
\begin{align}
    \textcolor{black}{\tilde{T_u}} = \{\underbrace{\vphantom{p_1, p_2, ..., p_L,}p_1, p_2, ..., p_L,}_{\text{prefix}} \underbrace{\vphantom{p_1, p_2, ..., p_L,}t_1, t_2, ..., t_W,}_{\text{input text}} \underbrace{\vphantom{p_1, p_2, ..., p_L,}s}_{\textcolor{black}{\text{suffix}}}\},
\end{align}
where $P=\{p_1, p_2, ..., p_L\}$ refers to the \emph{prefix} that contains $L$ prepended virtual tokens, and $s$ refers to the \emph{suffix} that consists of one single appended virtual token.
\textcolor{black}{During model training, we freeze the parameters of LLMs and tailor them for the recommendation task by optimizing the trainable virtual tokens added at the prefix $P$ and suffix $s$, which greatly reduces the size of the parameters to be trained.}

Specifically, the prefix \textcolor{black}{$P$} containing $L$ virtual tokens is responsible for adapting  LLMs to the recommendation task with collaborative information. As proven by previous works \cite{DBLP:conf/acl/LiL20, DBLP:conf/emnlp/LesterAC21}, these virtual tokens can serve as placeholders that allow LLMs to capture task-specific information during fine-tuning. \textcolor{black}{In addition, we also use the prefix $P$ to integrate ID-based collaborative information into LLMs via the proposed M-Former (detailed in Section \ref{mformer}), which has proven useful for improving the recommendation results \cite{DBLP:journals/corr/abs-2403-06447, DBLP:conf/www/ZhuWGHL24}. }

\label{suffix}
In addition to the prefix $P$, we also append a special trainable virtual token $s$ to the end of the input text $T_u$, which is called the \emph{suffix}. Previous works \cite{DBLP:journals/corr/abs-2305-11700} have tried to perform average pooling on the token embeddings output by LLMs to obtain the user embedding. However, most generative LLMs are based on the masked attention mechanism, dictating that only the last token can observe the entire input. \textcolor{black}{Therefore, these works may introduce noise by performing average pooling on all output embeddings.} In this work, we utilize an appended trainable virtual token \textcolor{black}{$s$} to capture the information of the entire input $\textcolor{black}{\tilde{T_u}}$, whose output embedding \textcolor{black}{$\mathbf{h}_s$} can be taken as the user embedding \textcolor{black}{$\mathbf{h}_u$} for similarity comparison and next item prediction. Formally, the encoding process of LLMs can be represented as:
% The rationale for the use of the suffix is obvious.
\begin{align}
    \{\mathbf{h}_{p_1}, ..., \mathbf{h}_{t_1}, ..., {\mathbf{h}}_{s} \} = \text{LLM}(\{\mathbf{e}_{p_1}, ..., \mathbf{e}_{t_1}, ..., \mathbf{e}_{s} \}),
\end{align}
where $\mathbf{e}\in\mathbb{R}^{d}$ represents an input token embedding of the input $\tilde{T_u} = \{p_1, ..., t_1, ..., s\}$, $\mathbf{h}\in\mathbb{R}^{d}$ represents the corresponding output embedding, and $d$ represents the hidden size of the LLM. Through the trainable suffix \textcolor{black}{$s$}, we can effectively convert the output embedding of the LLM from the language space to the recommendation space. When taking the user interaction history or the single item as input of the LLM, we can directly take the output suffix embedding $\mathbf{h}_s$ as the user embedding \textcolor{black}{$\mathbf{h}_u$} or item embedding \textcolor{black}{$\mathbf{h}_i$} for further recommendation.

\subsubsection{Item Recommendation}
\label{rec-sim}
Given the obtained user embedding $\mathbf{h}_u\in\mathbb{R}^{d}$ and the item embedding $\mathbf{h}_i\in\mathbb{R}^{d}$ from the LLM, we can compute the similarity between them as follows:
\begin{align}
s(u, i) & = \cos(\mathbf{h}_{u},\mathbf{h}_{i})= \frac{\mathbf{h}_{u}^{\top} \mathbf{h}_{i}}{\left\|\mathbf{h}_{u}\right\| \cdot\left\|\mathbf{h}_{i}\right\|},
\label{cosine}
\end{align}
where ${s(u, i)}\in\mathbb{R}$ indicates the probability that the item $i$ will become the next item browsed by user $u$. To predict the next item, we iterate through each item $i$ in the item set $\mathcal{I}$, and select the item 
$\hat{i}$ with the highest score as the next item:
\begin{align}
\hat{i} & = \operatorname{argmax}_{i \in \mathcal{I}}\left(s(u, i)\right).
\end{align}

\subsection{M-Former based Collaborative Information Integration}
\label{mformer}
In the proposed LLM Bi-Tuning, we use the trainable prefix \textcolor{black}{$P$} and suffix \textcolor{black}{$s$} to adapt the LLM for recommendation. In order to achieve better recommendation results, we incorporate the collaborative information via the prefix \textcolor{black}{$P$}. Existing works have tried to use unified linear layers to project the collaborative embeddings encoded by ID-based sequential recommender systems into the language space of LLMs \cite{DBLP:journals/corr/abs-2310-20487, DBLP:journals/corr/abs-2310-19488}. However, this method is too simple to detect the diverse characteristics of different types of users  \cite{DBLP:journals/isci/ZanZMLD21,DBLP:conf/www/LiCFGZ21}.  To deal with this challenge, we introduce M-Former, an MoE-based querying transformer that employs a set of query experts to deal with different types of users and integrates user-specific collaborative information into the prefix \textcolor{black}{$P$}.

% for better LLM adaptation and more accurate recommendation. 
In the following, we first describe the MoE strategy, namely how to select the appropriate query expert based on the collaborative characteristics of a specific user from a set of experts. Then we explain 
\textcolor{black}{how the selected query expert interacts with the collaborative information encoded by frozen ID-based sequential recommender systems in the querying transformer. }

\subsubsection{Mixture of Experts}
As shown in Figure \ref{model},  there are $K$ query experts dealing with users of different types, \textcolor{black}{each of which contains $L$ trainable virtual tokens}. In order to select the most appropriate expert to deal with user-specific  collaborative information,
we set up a router to calculate the scores of different experts given the specific user. Formally, given the user interaction history ${S}_u=\{i_1, i_2,\dots,i_{N}\}$ of user $u$, the corresponding item ID sequence ${ID}_u=\{{id}_{i_1}, {id}_{i_2},\dots,{id}_{i_{N}}\}$ is taken as input of a pre-trained ID-based sequential recommender system (frozen) and encoded as $\mathbf{C}_u\in\mathbb{R}^{N \times d_i}$, where 
$N$ represents the length of the interaction history $S_u$ and $d_i$ represents the hidden size of the sequential recommender system. Then,  $\mathbf{C}_u$ is sent into the router, which is implemented with a fully-connected layer in this work. The \textcolor{black}{matching degree} of the $K$ query experts according to the $N$ user-interactive items embedded as $\mathbf{C}_u$ can be calculated as:
\begin{equation}\textcolor{black}{m}(u)=\mathbf{C}_u\cdot \mathbf{W}_r^{\top},\end{equation}
where  $\mathbf{W}_r\in\mathbb{R}^{K\times d_i}$ is the router's linear weight, and $\textcolor{black}{m}(u)\in\mathbb{R}^{N\times K}$. Then, he $i$-th item's score for the $j$-th expert can be calculated as:
\begin{equation}
\label{score_matrix}
{p}_{i,j}(u)=\frac{e^{\textcolor{black}{m}_{i,j}(u)}}{\sum_{z=1}^Ke^{\textcolor{black}{m}_{i,z}(u)}},
\end{equation}
and the final score of the $j$-th query expert can be obtained by:
\begin{equation}
r_{j}(u) = \frac{\sum_{i=1}^N {p_{i,j}(u)}}{N}.
\end{equation}
Finally, the query expert with the highest score will be selected to deal with the specific user $u$.

\subsubsection{MoE-based Querying Transformer}

% \subsubsection{/}
\textcolor{black}{As described above, we obtain the most appropriate query expert to handle the specific user's collaborative information $\mathbf{C}_u$. Afterward, as shown in Figure \ref{model}, the selected query expert containing $L$ virtual tokens is sequentially fed into $Z$ transformer blocks to interact with the collaborative information $\mathbf{C}_u$. In this way, the query expert integrates the collaborative information into its $L$ trainable virtual tokens, which further act as the aforementioned prefix $P=\{p_1, p_2, ..., p_L\}$ to adapt LLMs for the sequential recommendation. 
} 

\textcolor{black}{Formally, the query expert can be represented as $\mathbf{E}\in\mathbb{R}^{L\times d_m}$, where $d_m$ is the hidden size of the M-Former.
In the transformer block,  $\mathbf{E}$ is first encoded by a self-attention layer and then projected to the query matrix $\mathbf{Q}$ used in the cross-attention layer to interact with the key/value matrix ($\mathbf{K}$/$\mathbf{V}$)  \textcolor{black}{projected} from the  ID-based collaborative embeddings $\mathbf{C}_u$. In this way, we update the query expert's embedding \textcolor{black}{$\mathbf{E}$} and integrate ID-based collaborative information into it. Then, through a linear projection layer set up on the top of the \textcolor{black}{$Z$} transformer blocks, the query expert is projected into the LLM's hidden size $d$ and acts as the prefix $P=\{p_1, p_2, ..., p_L\}$ to adapt the LLM for sequential recommendation with the enhancement of collaborative information.}

\subsection{Model Learning}

In this work, we employ a multi-task training strategy to train our Laser, which takes into account both the recommendation goal and the load balancing goal of the MoE experts. Furthermore,  we perform a two-stage training by first finding the most appropriate parameter weights to obtain high-quality item embeddings used for user-item similarity comparison, and then further training Laser based on these fixed item embeddings to achieve the best recommendation results.

\subsubsection{Loss Function}
We propose a multi-task training strategy to jointly train the proposed Laser for LLM-based sequential recommendation.

The first training task is the item-item contrastive (IIC) task, which is widely employed for next item prediction. Following previous work \cite{DBLP:conf/kdd/LiWLFSSM23}, we use the ground-truth next item as the positive instance and all other items in the item set $\mathcal{I}$ as negative instances. Formally, the item-item contrastive loss is calculated as:
\begin{equation}\mathcal{L}_{\mathrm{IIC}}=-\log\frac{e^{\cos(\mathbf{h}_u,\mathbf{h}_i^+)/\tau}}{\sum_{i\in\mathcal{I}}e^{\cos(\mathbf{h}_u,\mathbf{h}_i)/\tau}},\end{equation}
where the calculation of $\cos(\mathbf{h}_u,\mathbf{h}_i)$ is consistent with Equation (\ref{cosine}), $\mathbf{h}_i^+$ represents the embedding of the ground-trouth next item,  and $\tau$ is a temperature hyper-parameter.

The second training \textcolor{black}{task} is the load balancing task, which is used to encourage a balanced load across different query experts of the M-Former. As proven by previous works \cite{DBLP:journals/jmlr/FedusZS22, DBLP:conf/iclr/LepikhinLXCFHKS21}, this task can force the router to assign users with diverse collaborative characteristics to different query experts, such that each expert can be trained to obtain the best collaborative information integration effectiveness for its group of users. Formally, the load balancing loss is calculated as:
\begin{equation}\mathcal{L}_{\mathrm{LB}}= K\cdot\sum_{j=1}^Kf_j\cdot P_j,\label{lb}\end{equation}
where $K$ is the number of query experts, $f_j$ represents the fraction of items dispatched to the $j$-th expert that can be calculated as:
\begin{equation}
f_j=\frac{1}{N}\sum_{i=1}^{N}{\mathbbm{1}}\{\text{argmax }p_i(u)=j\},
\end{equation}
where $N$ is the number of items in the user interaction history, $p(u)\in\mathbb{R}^{N\times K}$ is the score matrix obtained through Equation (\ref{score_matrix}), which represents the degree of correlation between the $N$ items and the $K$ query experts. The  $P_j$ in Equation (\ref{lb}) represents the fraction of the router probability allocated for the $j$-th expert, which can be calculated as:
\begin{equation}
P_j=\frac1N\sum_{i=1}^{N}p_{i,j}(u).
\end{equation}
Totally, the loss function we use in this work is:
\begin{equation}
\mathcal{L}=\mathcal{L}_{\mathrm{IIC}}+\lambda\cdot\mathcal{L}_{\mathrm{LB}},\end{equation}
where $\lambda$ is a hyper-parameter that controls the weight of different tasks.

\subsubsection{Two-Stage Training}
\label{2-stage}
In this work, the recommendation is conducted based on the user-item embedding similarity comparison. Since the item embedding is determined by the corresponding trainable suffix, which changes after different training epochs. We employ a two-stage training strategy to first find the most appropriate parameter weights to obtain high-quality item embeddings, and then further train Laser based on these fixed item embeddings to achieve the best recommendation results.
 
Specifically, in the first training stage, at the beginning of each epoch, the item embeddings are updated as $\mathbf{I}\in\mathbb{R}^{|\mathcal{I}|\times d}$ using the current parameter weights $A$. Then, Laser is trained on  $\mathbf{I}$  and validated at the end of the epoch based on the updated parameter weights $A'$.
At the end of the first training stage, the embeddings  $\hat{\mathbf{I}}$ and the corresponding parameter weights $\hat{A'}$ of the best-performing epoch are selected for the second training stage. In the second training stage, Laser is initialized with $\hat{A'}$ and then trained for multiple epochs to further adapt to the fixed embeddings $\hat{\mathbf{I}}$. Finally, the parameter weights $\hat{A}$ that yield the optimal validation performance are reserved, and the test results performed on $\hat{A}$ and $\hat{\mathbf{I}}$ represent the final performance of Laser.

\begin{table}[t!]
% \small
\centering
\caption{Statistics of the preprocessed datasets. Avg. n denotes the average number of items in the user interaction history.}
% \vspace{-0.1in}
\label{tab:stat}
\scalebox{0.9}{
\setlength{\tabcolsep}{1mm}{
\begin{tabular}{lrrrrr}
\toprule
\multicolumn{1}{c}{\textbf{Datasets}} & \multicolumn{1}{c}{\textbf{\#Users}} & \multicolumn{1}{c}{\textbf{\#Items}} & \multicolumn{1}{c}{\textbf{\#Inters.}} & \multicolumn{1}{c}{\textbf{Avg. n}} & \multicolumn{1}{c}{\textbf{Density}} \\ \midrule
\textbf{Scientific}                   & 11,041                               & 5,327                                & 76,896                                 & 6.96                                & 1.3e-3                               \\
\textbf{Arts}                         & 56,210                               & 22,855                               & 492,492                                & 8.76                                & 3.8e-4                               \\
\textbf{Pet}                          & 47,569                               & 37,970                               & 420,662                                & 8.84                                & 2.3e-4                               \\ \bottomrule
\end{tabular}
}}
% \vspace{-0.1in}
\end{table}

\begin{table*}[t!]
\small
\centering
% \small
\caption{Performance comparison of different methods. The best results are in {bold}
and the second best results are {underlined}. Improv.~ indicates the improvement between the best and second best results.}
% \vspace{-3mm}
\label{tab:perf}
\scalebox{1.0}{
\setlength{\tabcolsep}{1mm}{
\begin{tabular}{llccccccccccc}
\toprule
          &       & \multicolumn{6}{c}{\textbf{Traditional Methods}} & \multicolumn{4}{c}{\textbf{LLM-based Methods}} &  \\
          \cmidrule(lr){3-8} \cmidrule(lr){9-12} 
    \textbf{Dataset} & \textbf{Metric} & \textbf{SASRec} & \textbf{BERT4Rec} & \textbf{RecGURU} & \textbf{ZESRec} & \textbf{RECFORMER} & \textbf{FDSA} & \textbf{LLM4REC} & \textbf{KAR} & \textbf{LlamaRec} & \textbf{Laser} & \textbf{Improv.} \\
    \midrule
    \multirow{3}[0]{*}{Scientific} & Recall@10 & \underline{0.1305}  & 0.1061  & 0.0781  & 0.1260  & 0.1114  & 0.0967  & 0.1257  & 0.1265  & 0.1275  & \textbf{0.1396 } & \textbf{6.97\%} \\
          & NDCG@10 & 0.0797  & 0.0790  & 0.0575  & 0.0843  & 0.0722  & 0.0716  & 0.0764  & \underline{0.0894}  & 0.0857  & \textbf{0.0970 } & \textbf{8.54\%} \\
          & MRR   & 0.0696  & 0.0759  & 0.0566  & 0.0745  & 0.0650  & 0.0692  & 0.0683  & \underline{0.0813}  & 0.0793  & \textbf{0.0893 } & \textbf{9.81\%} \\ \midrule
    \multirow{3}[0]{*}{Pet} & Recall@10 & 0.0881  & 0.0765  & 0.0415  & \underline{0.1018}  & 0.0905  & 0.0949  & 0.0918  & 0.0942  & 0.0961  & \textbf{0.1134 } & \textbf{11.37\%} \\
          & NDCG@10 & 0.0569  & 0.0602  & 0.0366  & 0.0754  & \underline{0.0793}  & 0.0673  & 0.0769  & 0.0724  & 0.0754  & \textbf{0.0898 } & \textbf{13.27\%} \\
          & MRR   & 0.0507  & 0.0585  & 0.0371  & 0.0706  & \underline{0.0774}  & 0.0650  & 0.0681  & 0.0677  & 0.0711  & \textbf{0.0856 } & \textbf{10.63\%} \\\midrule
    \multirow{3}[0]{*}{Arts} & Recall@10 & 0.1342  & 0.1236  & 0.0742  & 0.1349  & 0.1298  & 0.1209  & 0.1266  & 0.1357  & \underline{0.1368}  & \textbf{0.1489 } & \textbf{8.91\%} \\
          & NDCG@10 & 0.0848  & 0.0942  & 0.0525  & 0.0970  & \underline{0.1024}  & 0.0994  & 0.0927  & 0.0917  & 0.0860  & \textbf{0.1138 } & \textbf{11.17\%} \\
          & MRR   & 0.0742  & 0.0899  & 0.0488  & 0.0870  & \underline{0.0980}  & 0.0941  & 0.0880  & 0.0818  & 0.0794  & \textbf{0.1095 } & \textbf{11.80\%} \\ \bottomrule
\end{tabular}
}}
\end{table*}

\section{Experiments}
In this section, we conduct detailed experiments to demonstrate the effectiveness of our proposed Laser.

\subsection{Experimental Setup}
\subsubsection{Datasets}
To evaluate the effectiveness of our Laser, we conduct experiments on three categories of the Amazon review datasets \cite{DBLP:conf/emnlp/NiLM19}, including “Industrial and Scientific”, “Arts, Crafts and Sewing”, and “Pet Supplies”. Following previous works \cite{DBLP:conf/kdd/LiWLFSSM23, DBLP:conf/kdd/HouMZLDW22}, we use the five-core datasets provided by the data source and filter out items with missing titles. Then, we collect the interactions for different users and sort the interactive items by timestamp in ascending order. The statistics of the preprocessed datasets are shown in Table \ref{tab:stat}. As for the item semantic information modeling, we select the item attributes including title, category, and brand.

\subsubsection{Baselines and Implementation Details} \label{implement}
We compare our Laser to a number of state-of-the-art baselines, including six traditional methods  (\textbf{SASRec}~\cite{DBLP:conf/icdm/KangM18}, \textbf{BERT4Rec}~\cite{DBLP:conf/cikm/SunLWPLOJ19}, \textbf{RecGURU}~\cite{DBLP:conf/wsdm/LiZZYCSKN22}, \textbf{FDSA}~\cite{DBLP:conf/ijcai/ZhangZLSXWLZ19}, \textbf{ZESRec}~\cite{DBLP:journals/corr/abs-2105-08318}, \textbf{RECFORMER}~\cite{DBLP:conf/kdd/LiWLFSSM23}), and three LLM-based methods (\textbf{LLM4REC}~\cite{DBLP:journals/corr/abs-2402-09617}, \textbf{ZESRec}~\cite{DBLP:journals/corr/abs-2105-08318}, \textbf{LlamaRec}~\cite{DBLP:journals/corr/abs-2311-02089}). We list the details of these baselines in the Appendix \ref{app:baselines}. Besides, in this paper, the frozen ID-based sequential recommender employed in the Laser is a pre-trained BERT4Rec \cite{DBLP:conf/cikm/SunLWPLOJ19}, and the utilized frozen LLM is the ChatGLM2-6B \cite{glm2024chatglm}. The other trainable modules are all randomly initialized. The settings of each module and other implementation details are shown in Appendix \ref{app:implement}.

\subsubsection{Evaluation Settings}
Following previous works \cite{DBLP:conf/kdd/LiWLFSSM23, DBLP:journals/corr/abs-2402-09617, DBLP:journals/corr/abs-2311-02089}, we employ three popular metrics, including Recall@N, NDCG@N and MRR, where N is set to 10. For data splitting, we adopt the leave-one-out \cite{DBLP:conf/icdm/KangM18} strategy, where the most recent item in the interaction history is used for testing, the second most recent item is used for validation, and the remaining items are used for training. We treat all items in the item set as candidate items and report the average results on the test data.

\subsection{Overall Performance}

As shown in Table \ref{tab:perf}, we compare our proposed Laser to nine state-of-the-art baselines across three Amazon datasets. From the experimental results, we can obtain following observations.

First, compared to the other outstanding sequential recommendation methods, our Laser results in significant improvements on all metrics across all datasets. For example, on the Pet dataset, compared to the second best method, our Laser improves Recall@10, NDCG@10, and MRR by around 11.37\%, 13.27\%, and 10.63\%, respectively. 
% In the Arts dataset, these three metrics are increased by 8.91\%, 11.17\%, and 11.8\%, respectively. 
This demonstrates that our proposed framework can successively adapt LLMs to effective sequential recommender systems. We believe that our Laser benefits from the Bi-Tuning method that effectively adapts LLMs for sequential recommendation with collaborative information. In addition, when integrating the collaborative information, the designed M-Former (MoE-based querying transformer)  captures
the diverse characteristics of different types of users for more accurate recommendation.

\begin{table}[t!]
\centering
% \small
\caption{Parameter scale comparison of different LLM tuning methods.}
% \vspace{-3mm}
\label{tab:lmm}
\scalebox{0.9}{
\setlength{\tabcolsep}{1mm}{
\begin{tabular}{lccc}
\toprule
    \multirow{2}{*}{\textbf{Method}} & \multirow{2}{*}{\textbf{Backbone}} & \textbf{Tuning} & {\textbf{Trainable}} \\  & &\textbf{Method} & \textbf{Parameters}\\ \midrule
{LLM4REC} & GPT2-Large & Full Fine-Tuning & 787.3M \\
{KAR} & ChatGPT & /     & / \\
{LlamaRec} & Llama2-7B & QLoRA & 4.194M \\
{Laser} & ChatGLM2-6B & Bi-Tuning & 0.135M \\
\bottomrule
\end{tabular}
}}
\vspace{-0.1in}
\end{table}

Second, compared to the traditional methods, the three LLM-based baselines do not always yield better results. A possible reason is that LLMs have not been pre-trained on large amounts of recommendation data, resulting in lacking the task-specific knowledge \cite{DBLP:journals/corr/abs-2308-08434, DBLP:conf/recsys/BaoZZWF023}. This further illustrates the importance of exploring more appropriate methods to adapt LLMs to recommendation tasks more effectively. In contrast, our Laser significantly outperforms all traditional methods on all metrics across all datasets. For example, on the Scientific dataset, compared to the best traditional method, our Laser improves Recall@10, NDCG@10, and MRR by 6.97\%, 15.07\%, and 17.65\%, respectively, demonstrating the validity of our method and hopefully inspiring future LLM-based recommendation works.

% Table generated by Excel2LaTeX from sheet 'Sheet4'
\begin{table*}[t!]
% \small
  \centering
  \caption{Results of the ablation study. The best results are in {bold}
and the second best results are {underlined}.}
% \vspace{-3mm}
\scalebox{0.9}{
\setlength{\tabcolsep}{1mm}{
    \begin{tabular}{lcccccc}
    \toprule
   & \multicolumn{3}{c}{\textbf{Scientific}} & \multicolumn{3}{c}{\textbf{Pet}} \\
          \cmidrule(lr){2-4} \cmidrule(lr){5-7}
           {\textbf{Variants}} & \textbf{Recall@10} & \textbf{NDCG@10} & \textbf{MRR}   & \textbf{Recall@10} & \textbf{NDCG@10} & \textbf{MRR} \\ 
          \midrule
    Laser & \textbf{0.1396 } & \textbf{0.0970 } & \textbf{0.0893 } & \textbf{0.1134 } & \textbf{0.0898 } & \textbf{0.0856 } \\ 
    w/o MoE & 0.1261  & 0.0889  & \underline{0.0795}  & 0.1056  & 0.0818  & 0.0763  \\
    w/o M-Former & 0.1245  & 0.0844  & 0.0739  & 0.1049  & 0.0775  & 0.0721  \\
    w/o prefix & 0.1128  & 0.0705  & 0.0619  & 0.0878  & 0.0695  & 0.0607  \\
    w/o training stage 1 & 0.0784  & 0.0544  & 0.0509  & 0.0514  & 0.0443  & 0.0398  \\
    w/o training stage 2 & \underline{0.1316}  & \underline{0.0894}  & 0.0781  & \underline{0.1083}  & \underline{0.0837}  & \underline{0.0793}  \\
    \bottomrule
    \end{tabular}%
    }}
      % \vspace{-1mm}
  \label{ablation}%
\end{table*}%

Additionally, we compare the parameter scale of different LLM tuning methods in Table \ref{tab:lmm}. We can observe that our proposed Bi-Tuning is more parameter-efficient, which contains only 0.135M trainable parameters.
% for LLM adaptation. 
% to the sequential recommendation task. 
The results indicate that Laser can greatly reduce the scale of trainable parameters and achieve effective LLM adaptation with the proposed Bi-Tuning method, outperforming SOTA baselines. 
Note that in Table \ref{tab:lmm} we only list the trainable parameter scales of different LLM tuning methods. The total number of Laser's trainable parameters is about 183.3M, which is also significantly less than 3\% of the parameter number of the LLM backbone, ChatGLM2-6B.

\begin{figure}[t!] %H为当前位置，!htb为忽略美学标准，htbp为浮动图形
    \centering %图片居中
    \includegraphics[width=0.45\textwidth]{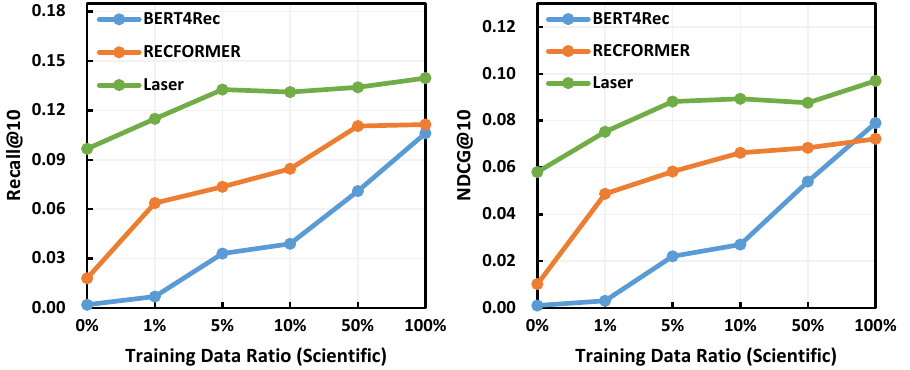} %插入图片，[]中设置图片大小，{}中是图片文件名
    % \vspace{-0.1in}
    \caption{\textcolor{black}{Performance comparison under the zero-shot and low-resource settings on the Scientific dataset.}} %最终文档中希望显示的图片标题
    \label{low-resource} %用于文内引用的标签
    % \vspace{-0.1in}
\end{figure}%结束环境

\subsection{{Zero-Shot and Low-Resource Performance}}

To further demonstrate the effectiveness of our Laser, we perform experiments to examine its performance in the zero-shot and low-resource scenarios. Specifically, we compare Laser (which uses both the item semantic information and the ID-based collaborative information) with two other types of baselines, including BERT4Rec (which uses only the ID-based collaborative information) and RECFORMER (which uses only the item semantic information). We first pre-train these methods (in addition to the ID-based BERT4Rec) on the Pet dataset, and then test whether they can perform well on another domain with no/limited training data. 

Figure \ref{low-resource} shows the experimental results on the Scientific dataset. From Figure \ref{low-resource}, we can observe that: (1) Laser performs best in the zero-shot scenario. Compared to other baselines, Laser achieves significantly better performance (Recall@10 reaches 0.97, NDCG@10 reaches 0.58), even though it has not seen any items on the Scientific dataset.  We attribute this superior performance to the design of our Bi-Tuning framework, which fully leverages the generalization capabilities of LLMs and effectively adapts LLMs for sequential recommendation. 
(2) Laser only needs to use 5\% of the training data to exceed the effect of the other two baselines using 100\% of the training data. Compared to the other two baselines, Laser's performance can quickly rise to a very appreciable level as the ratio of training data increases to 5\%. This means that we only need a very small amount of training data and training time to migrate the Laser trained on one domain to another unseen domain, accompanied by better results than other baselines that need far more training data. This demonstrates that our proposed framework can effectively transform LLMs into generalizable sequential recommender systems.

\begin{table}[t!]
% \small
  \centering
  \caption{Performance comparison of different item/user embedding generation strategies on the Scientific dataset. The best results are in {bold}
and the second best results are {underlined}.}
\vspace{-1mm}
\scalebox{0.9}{
\setlength{\tabcolsep}{1mm}{
    \begin{tabular}{lcccccc}
    \toprule

\textbf{Strategies} & \textbf{Recall@10} & \textbf{NDCG@10} & \textbf{MRR} \\ \midrule
    {w/ suffix} & \textbf{0.1396} & \textbf{0.0970} & \textbf{0.0893} \\
    {w/ average pooling} & 0.0551  & 0.0416  & 0.0401  \\
    {w/ [EOS]} & \underline{0.0948}  & \underline{0.0688}  & \underline{0.0646}  \\
    \bottomrule
    \end{tabular}%
    }}
  \label{embedding_generation}%
      \vspace{-2mm}
\end{table}%

\subsection{Ablation Study}

To demonstrate the effectiveness of each module in our Laser, we conduct ablation studies and provide the results in Table \ref{ablation}. 
We can observe that: (1) The experimental results on the two datasets remain identical. The removal of any module results in a significant decrease in Laser's performance. (2) Without MoE, Recall@10, NDCG@10, and MRR  respectively decrease on average by 8.28\%, 8.63\%, and 10.92\%, showing that the introduction of MoE can help our framework to deal with the diverse collaborative characteristics of different types of users, which
% in a more detailed way, 
leads to higher-quality recommendation results. Furthermore, without the M-Former, the three metrics decrease on average by 9.15\%, 13.35\%, and 16.51\%, respectively.
% , across both datasets. 
This demonstrates the importance of using ID-based collaborative information for more accurate recommendations, and that our M-Former can effectively integrate collaborative information into LLMs.  (3) Without the prefix, Recall@10, NDCG@10, and MRR decrease significantly on average by 20.89\%, 24.97\%, and 25.39\%, respectively, demonstrating the important role of the prefix in adapting LLMs to the recommendation task. (4) Removing any training stage, the effectiveness of the Laser is reduced. Specifically, without the first training stage, Recall@10, NDCG@10, and MRR respectively decline on average by 49.24\%, 47.29\%, and  48.28\%, demonstrating the need to find appropriate parameter weights to obtain the high-quality item embeddings. Besides, without the second training stage, the metrics also decrease by 9.13\%, 7.28\%, and 9.96\%, respectively. This suggests that after obtaining appropriate item embeddings, it's also necessary to continue training to make Laser better adapt to the fixed item embeddings and achieve the best recommendation results. 

% Table generated by Excel2LaTeX from sheet 'Sheet4'
\begin{table}[t]
% \small
  \centering
  \caption{Performance comparison under different hard prompt templates on the Scientific dataset. The best results are in bold and the second best results are underlined.}
  \vspace{-1mm}
  \scalebox{0.9}{
\setlength{\tabcolsep}{1mm}{
    \begin{tabular}{lccc}
    \toprule
       % &  \multicolumn{3}{c}{\textbf{Scientific}} \\
       %    \cmidrule(lr){2-4} 
        \textbf{Templates}  & \textbf{Recall@10} & \textbf{NDCG@10} & \textbf{MRR} \\ \midrule
    original & \textbf{0.1396 } & \textbf{0.0970 } & \textbf{0.0893 } \\  
    w/o specified phrase & \underline{0.1042}  & \underline{0.0814}  & \underline{0.0782}  \\
    w/o instruction & 0.0985  & 0.0747  & 0.0652  \\
    w/\;two instructions & 0.0972  & 0.0635  & 0.0567  \\
    \bottomrule
    \end{tabular}%
    }}
      \vspace{-2mm}
  \label{prompt exp}%
\end{table}%

\subsection{Further Discussion}
In this section, we provide further discussion about our proposed Laser.

\subsubsection{Suffix}
We compare the usage of the suffix to two other strategies for generating user/item embeddings, including performing average pooling on all token embeddings output by LLMs and replacing the trainable virtual suffix with a hard token [EOS] which will not be trained. As shown in table \ref{embedding_generation}, compared to the other two strategies, our designed suffix can effectively improve Recall@10, NDCG@10, and MRR by at least 32.07\%, 29.04\%, and 27.67\%, respectively. This proves that a trainable virtual suffix can more effectively convert the LLM output from the language space to the recommendation space, thus generating higher-quality user/item embeddings.

\subsubsection{Input Text Template}
\label{prompt_template_exp}
In this work, we utilize a unified
 template to organize both the user interaction history and the single item, which is shown in Section \ref{prompt}. The template instructs LLMs to summarize the semantic information into the suffix, which is further used for recommendation. To ensure the template's plausibility, we compared it with three other variants, including: (1) deleting the specified instruction phrase ``into a single token'', (2) deleting the entire instruction ``You are an intelligent ... the user has browsed the following items:'', (3) using a different instruction for item embedding generation, ``You are an intelligent recommendation assistant. Please summarize the item characteristics into a single token:''. As shown in Table \ref{prompt exp}, compared to the other three variants, our prompt template can significantly improve Recall@10, NDCG@10, and MRR by at least 25.36\%, 16.05\%, and 12.36\%, respectively. This demonstrates the effectiveness of our prompt template in harnessing the powerful capabilities of LLMs with clear, consistent, and appropriate instruction.

\subsubsection{ID-based Sequential Recommender.}

\begin{figure}[t!] %H为当前位置，!htb为忽略美学标准，htbp为浮动图形
    \centering %图片居中
    \includegraphics[width=0.40\textwidth]{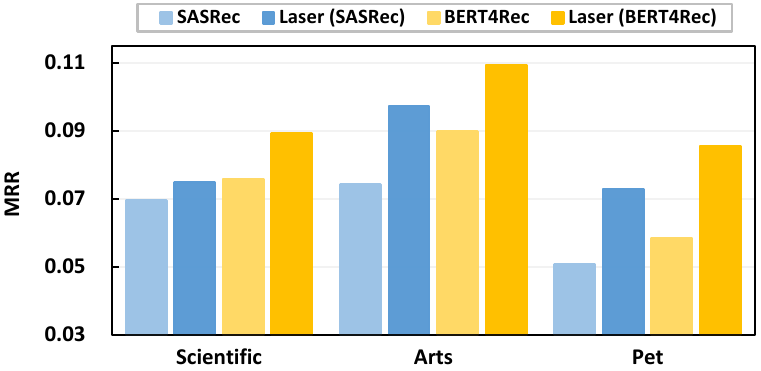} %插入图片，[]中设置图片大小，{}中是图片文件名
      \vspace{-2.5mm}
    \caption{\textcolor{black}{Performance comparison with different ID-based sequential recommender systems.}} %最终文档中希望显示的图片标题
    \label{seqrec} %用于文内引用的标签
    \vspace{-2mm}
\end{figure}

We perform further experiments to study the effect of the ID-based sequential recommender system on the performance of our Laser. 
As shown in Figure \ref{seqrec}, on all datasets, the performance of Laser increases almost linearly with the performance of the employed ID-based sequential recommender system. For example, on the Pet dataset, Laser based on BERT4Rec outperforms Laser based on SASRec by 17.30\%, while  BERT4Rec outperforms SASRec by 15.38\%. This suggests that our Laser can be further improved by using more powerful ID-based sequential recommender systems. 

\subsubsection{Parameter Analysis.}
Furthermore, we perform a detailed \textcolor{black}{parameter analysis} to explore the effect of the query expert number $K$ and the expert's virtual token number $L$ on the Laser's performance. As shown in Table \ref{parameter}, smaller values of $K$ and $L$ cause the M-Former to be insufficient to effectively handle the diverse characteristics of different types of users, while too large values increase the difficulty of training, thus decreasing the effectiveness of Laser. Finally, we respectively set $K$ and $L$ to 8 and 32 to get the best results.

%结束环境

\section{Related Work}
\subsection{Sequential Recommendation}
Sequential recommendation aims to infer users' preferences based on their past interactions ordered by timestamps. 
In traditional sequential recommender systems, items are represented by unique IDs. To effectively capture users' historical interactions and make recommendations based on these IDs, a variety of methods 
have been employed, including CNNs, RNNs, and GNNs. 
For example, Caser \cite{DBLP:conf/wsdm/TangW18} views the embedding matrix of previous items as an "image" and applies convolutional operations to capture user preferences. GRU4Rec \cite{DBLP:journals/corr/ChungGCB14} introduces GRU \cite{DBLP:journals/corr/ChungGCB14} to model user sequential patterns.
SRGNN \cite{DBLP:conf/aaai/WuT0WXT19}, GCE-GNN \cite{DBLP:conf/sigir/Wang0CLMQ20}, and SURGE \cite{DBLP:conf/sigir/ChangGZHNSJ021} are proposed to capture long-term sequential user preferences through multi-layer message passing.  
 Besides, self-attention-based models have also been widely adopted for sequential recommendation \cite{DBLP:conf/icdm/KangM18, DBLP:conf/wsdm/LiWM20, DBLP:conf/cikm/SunLWPLOJ19}. Although these ID-based methods achieve promising performance, they fail to consider the semantic information of item descriptions, resulting in suboptimal performance.
Recently, researchers have attempted to 
create transferable item representations by encoding item descriptions with language models \cite{DBLP:conf/kdd/HouMZLDW22, DBLP:conf/kdd/LiWLFSSM23}. However, these works primarily focus on small or medium-sized language models.

\subsection{LLMs in Recommender Systems}
Large Language Models (LLMs) have demonstrated remarkable performance in various domains, prompting researchers to explore their potential in recommendation tasks. Existing works integrate LLMs into recommendations in two main paradigms. The first paradigm is to use LLMs to answer specific recommendation questions by in-context learning \cite{DBLP:conf/cikm/HeXJSLFMKM23, DBLP:conf/ecir/HouZLLXMZ24, DBLP:journals/corr/abs-2304-10149, DBLP:conf/recsys/ZhangBZWF023} or supervised fine-tuning \cite{DBLP:journals/corr/abs-2308-08434, DBLP:conf/recsys/BaoZZWF023, DBLP:conf/www/LinSZDCQTY024, DBLP:conf/aaai/WuQZZC24, DBLP:journals/corr/abs-2305-07001},
which focuses only on the reranking phase. The second paradigm is to use LLMs as encoders to generate item/user embeddings. For example, \citet{DBLP:journals/corr/abs-2305-11700} and \citet{DBLP:conf/sigir/WuWQ021} attempted to get item/user embeddings by performing pooling on the token embeddings encoded by LLMs. \citet{ DBLP:journals/corr/abs-2403-01744} compressed the textual information into a single special token and learned its embedding using a contrastive learning approach. Although these works show promise,
% for extracting textual information into a single embedding that can be used for next token prediction,
they often require training a large number of parameters to bridge the huge gap between recommendation and language generation tasks, which is resource-demanding. Additionally, these works typically integrate ID-based collaborative signals into LLMs via simple unified linear projections \cite{DBLP:journals/corr/abs-2310-20487, DBLP:journals/corr/abs-2310-19488}, which fails to consider the diverse characteristics of different types of users and thus
diminishes recommendation accuracy.
In this paper, we propose a parameter-efficient LLM Bi-Tuning framework for sequential recommendation. Besides, to improve the recommendation performance, we introduce a lightweight M-Former to effectively integrate ID-based collaborative information into the LLM.

\begin{table}[t!]
% \small
  \centering
  \caption{The comparison under different $K$ and $L$ values on the validation set of the Scientific dataset. The best results are in bold and the second best results are underlined.}
  \vspace{-3mm}
  \scalebox{0.9}{
\setlength{\tabcolsep}{1mm}{
    \begin{tabular}{llccc}
    \toprule
       % & & \multicolumn{3}{c}{\textbf{Scientific}} \\
       %    \cmidrule(lr){3-5} 
     \textbf{K} & \textbf{L}    & \textbf{Recall@10} & \textbf{NDCG@10} & \textbf{MRR} \\ \midrule
    8    &32 & \textbf{0.1661 } & \textbf{0.1199 } & \textbf{0.1112 } \\ \midrule
    4    &32 & 0.1545  & 0.1090  & 0.0992  \\
    12  &32 & \underline{0.1560}  & \underline{0.1138}  & \underline{0.1054}  \\  \midrule 
    8&    16 & 0.1487  & 0.1099  & 0.1032  \\
    8 &   48 & 0.0975  & 0.0692  & 0.0644  \\ \bottomrule
    \end{tabular}%
    }}
  \label{parameter}%
   \vspace{-2mm}
\end{table}%

\section{{Conclusion}} 
In this paper, we propose Laser, a parameter-efficient LLM Bi-Tuning framework for sequential recommendation with collaborative information. Specifically,  we present Bi-Tuning, a parameter-efficient fine-tuning method that adapts LLMs to sequential recommendation through the trainable prefix and suffix. The prefix adapts LLMs to the recommendation task with collaborative information, while the suffix converts LLM output from the language space to the recommendation space and obtains high-quality user/item embeddings. To effectively integrate ID-based collaborative information for more accurate recommendation, we introduce M-Former, a lightweight MoE-based querying transformer that uses a set of query experts to capture the diverse collaborative characteristics of different types of users. Finally, a multi-task loss function and a two-stage training strategy are employed to train Laser for the sequential recommendation. Extensive experiments on real-world datasets demonstrate that Laser can parameter-efficiently adapt LLMs to effective recommender systems, significantly outperforming state-of-the-art methods.

\bibliographystyle{ACM-Reference-Format}
\newpage
\bibliography{sample-base}

% \newpage
\appendix

\section{Baselines}
\label{app:baselines}

To comprehensively evaluate the performance of
our proposed Laser, we compare it to state-of-the-art baselines, including six traditional methods and three LLM-based methods.

(1) Traditional Baselines:

\begin{itemize}
    \item \textbf{SASRec}~\cite{DBLP:conf/icdm/KangM18} employs a self-attention mechanism to capture the semantic relevance between the user interaction sequence and the candidate items.
    \item \textbf{BERT4Rec}~\cite{DBLP:conf/cikm/SunLWPLOJ19} is a bidirectional self-attentive model, employing the cloze objective  to model users’ dynamic preferences from their historical behaviors.
    \item \textbf{RecGURU}~\cite{DBLP:conf/wsdm/LiZZYCSKN22} introduces an adversarial learning method to incorporate user information across domains and obtain generalized user representations for sequential recommendation.
    \item \textbf{FDSA}~\cite{DBLP:conf/ijcai/ZhangZLSXWLZ19} proposes a feature-level self-attention network that integrates different heterogeneous features of items into feature sequences with different weights through a vanilla attention mechanism.
    \item \textbf{ZESRec}~\cite{DBLP:journals/corr/abs-2105-08318} utilizes a pre-trained language model to convert item descriptions into feature representations.
    \item \textbf{RECFORMER}~\cite{DBLP:conf/kdd/LiWLFSSM23} formulates items as key-value attribute pairs and utilizes pre-trained language models to encode them for ID-free sequential recommendation.
\end{itemize}

(2) LLM-based Baselines:
\begin{itemize}
    \item \textbf{LLM4REC}~\cite{DBLP:journals/corr/abs-2402-09617} proposes a graph knowledge guided attentive LLM recommendation backbone to inject graph edge information into LLMs.
    \item \textbf{KAR}~\cite{DBLP:journals/corr/abs-2306-10933} proposes a hybrid-expert adapter that condenses  LLM-generated world knowledge into augmented vectors to  enhance the performance of recommendation models.
    \item \textbf{LlamaRec}~\cite{DBLP:journals/corr/abs-2311-02089} adopts a verbalizer-based approach that transforms LLM output logits into probability distributions over the candidate items.
\end{itemize}

\section{Implementation Details}
\label{app:implement}

In this work, we employ a pre-trained BERT4Rec \cite{DBLP:conf/cikm/SunLWPLOJ19} as the frozen ID-based sequential recommender system to encode the user interaction ID sequences. The hyper-parameter settings keep the same as in the original paper, where the number of transformer blocks, the number of attention heads, and the dimension of each attention head are set to 2, 2, and 32, respectively.

The frozen LLM we use is the ChatGLM2-6B \cite{glm2024chatglm}, an impressive open-source large language model with exceptional language modeling capabilities. This model consists of 28 transformer blocks. The hidden size is set to 4096 and the number of attention heads is set to 32. In the feed-forward networks, the dimension of the intermediate layer is set to 13,696. Besides, the model vocabulary consists of 65,024 unique tokens.

As for the M-Former, it contains  12 transformer blocks, with alternate blocks conducting cross-attention
between the collaborative embeddings and the virtual query expert tokens. The hidden size and the number of attention heads are set to 768 and 12, respectively. The expert number $K$,  query token number $L$, and the hidden size of the virtual tokens are set to 8, 32, and 768, respectively. Besides, the router and the projection layer are all implemented by single fully-connected layers, whose input/output dimensions are set to 64/8 and 768/4096, respectively. 

During the training process, the BERT4Rec and the ChatGLM2 are frozen.  We randomly initialize the other trainable modules and train them for two stages (as described in Section \ref{2-stage}). {Specifically, we set the batch size to 4 and the learning rate to 1e-4.
The loss weight hyper-parameter $\lambda$ is set to 0.01, and the loss temperature hyper-parameter $\tau$ is set to 0.05. We use the
Adam optimizer and train Laser for 15/5 epochs in the first/second training stage, respectively.}

\end{document}